# "While a Storm is Raging on the Open Sea":

# Regional Development in a Knowledge-based Economy

Loet Leydesdorff

Science & Technology Dynamics, University of Amsterdam

Amsterdam School of Communications Research (ASCoR)

Kloveniersburgwal 48, 1012 CX  Amsterdam, The Netherlands

loet@leydesdorff.net ; http://www.leydesdorff.net

**Abstract**

The Triple Helix of university-industry-government relations is elaborated into a systemic model that accounts for interactions among three dimensions. By distinguishing between the respective micro-operations, this model enables us to recombine the "Mode 2" thesis of a new production of scientific knowledge and the study of systems of innovation with the neo-classical perspective on the dynamics of the market. The mutual information in three dimensions provides us with an indicator for the self-organization of the resulting network systems. The probabilistic entropy in this mutual information can be negative in knowledge-based configurations. The knowledge base of an economy can be considered as a second-order interaction effect among interactions at interfaces between institutions and functions in different spheres. Proximity enhances the chances for couplings and, therefore, the formation of technological trajectories. The next-order regime of the knowledge base, however, can be expected to remain pending as selection pressure.

**JEL Classifications**: O14, O30, R11, R58

## 1. Introduction

Since 1996, when the Organization for Economic Cooperation and Development (OECD) formulated the concept of a "knowledge-based economy" (Abramowitz & David, 1996; Foray & Lundvall, 1996; OECD, 1996), the search for indicators specifically suited for measuring the "knowledge base" of an economy has been a priority of this organization. Godin (2004) argued in a recent review (elsewhere in this issue) that the OECD mainly relabeled its existing indicators from "science and technology policies," as well as some indicators from the "output and impact" category, into the categories of "creation and diffusion of knowledge." The author drew the conclusion that "the knowledge-based economy is above all a rhetorical concept." In this paper I argue that the shift from innovation theory and innovation indicators towards the search for indicators of "knowledge-based systems" can be elaborated both theoretically and methodologically. The mutual information in three dimensions is proposed as an indicator for measuring the knowledge base of an economy.

## 2. Innovations, trajectories, niches

David & Foray (2002: 9) noted that the knowledge-based economy "marks a break in the continuity with earlier periods, more a 'sea-change' than a sharp discontinuity." Indeed, the knowledge-based economy can be considered as transforming "the ship while a storm is raging on the open sea" (Neurath *et al.*, 1929). The development of indicators, however, tends to lag behind the development of relevant policy-making and theorizing.

For example, the biannual series of *Science Indicators* was initiated by the U.S. National Science Board in 1972. In 1987 these reports were renamed *Science and Engineering Indicators* in order "to reflect an increased awareness of the complementary roles played by science and engineering research and education in creating both new knowledge and new technological products and processes" (National Science Board, 1987: ix). The shift from a science policy perspective to innovation policies, however, was made much earlier as a response to the oil crises of the 1970s (e.g., OECD, 1980; Rothwell & Zegveld, 1981). Innovation indicators were eventually developed by the OECD in the so-called *Oslo Manual* (OECD, 1992; OECD/Eurostat, 1997).

First, the linear models of innovation which either assumed "market pull" or "technology push" as the driving forces of innovation had to be superseded. These linear models induce a focus on either economic indicators or S&T indicators, respectively. The interface between these two dimensions is more difficult to indicate. However, innovation takes place at an interface, almost by definition (Mowery & Rosenberg, 1979).

Within economic theorizing the importance of innovation for economic growth was recognized already in the 1950s (Abramowitz, 1956; Solow, 1957). Knowledge production was then considered exogenous to the economic process. Nelson & Winter (1977) in a well-known article entitled "In search of useful theory of innovation" proposed to explain differences in growth rates among sectors of the economy in terms of the structural relations to technological trajectories. Sectors are differently affected by technological developments (Pavitt, 1984).

For example, the banking sector is restructured by information and communication technologies at a pace different from agriculture (Barras, 1990; Freeman & Perez, 1989). Furthermore, innovations can develop momentum along technological trajectories (Hughes, 1987). However, the development of an interface into a *system of innovations* can be expected to change the institutions which were interfaced during the process. Sahal (1985) defined such a recombination as a system's innovation. The two sides of the interface may thereafter become locked-in into each other as in a coevolution (Arthur, 1989). A new trajectory can be expected in this case.

While the market can be idealized as an open network seeking equilibrium, innovation first requires closure of the network in terms of relevant stakeholders (Callon, 1998). Innovations are generated and incubated by locally producing units such as scientific laboratories, artisan workshops, and communities of instrument makers, but in interaction with market forces. As a composed construct, therefore, innovation has both a market dimension and a systemic dimension. The two dimensions are traded off at interfaces: what can be produced in terms of technical characteristics versus what can be diffused on relevant markets in terms of service characteristics (Lancaster, 1979; Saviotti, 1996; Frenken, 2001).

At some places, but not at others the selective interactions can be stabilized. When a dominant design emerges (Abernathy & Utterback, 1978; Utterback & Suarez, 1993), the evolutionary landscape changes because of the steep learning curves following upon a breakthrough in the marketplace (Arrow, 1962; Rosenberg, 1982). Thus, a competitive edge can be shaped locally. Such a locally shielded network density can also be considered as a niche (Kemp *et al.*, 1998). Interface problems that may lead to transaction costs can be solved within niches more easily than in their surroundings (Williamson, 1985; Biggiero, 1998).

**3. From Interfaces to "Systems of innovation"**

Unlike organizations niches have no fixed delineations. They can be considered as densities of interfaces in an environment that is otherwise more loosely connected. The accumulation of interfaces allows for a competitive advantage by reducing transaction costs within the niche. A niche can be shaped, for example, within the context of a multinational and diversified corporation or, more generally, within the economy. Porter (1990), for example, proposed to analyze national economies in terms of clusters of innovation. Clusters may span industrial sectors, business columns, and horizontal integrations. They can be expected to act as systems of innovation that proceed more rapidly than their relevant environments and thus are able to maintain a competitive edge.

In other words, systems of innovations can be considered as complex systems because they are based on maintaining interfaces in a variety of dimensions. Sometimes, the geographical delineation of niches can be straightforward as in the case of Italian industrial districts. These comprise often only one or a few valleys (Beccatini *et al.*, 2003; Biggiero, 1998). For cultural or political reasons one may wish to define a system of innovation *a priori* as national or regional. However, an innovation system evolves, and its shape is therefore not given *ex ante*. One may entertain the *hypothesis* of an innovation system, but the operationalization and the measurement remain crucial.

Riba & Leydesdorff (2001), for example, could show that it was not possible to indicate a Catalonian system of innovations in terms of their definition of knowledge-intensive indicators despite the announcement of such a system prevailing in the literature on the basis of occupational and sectoral indicators (Braczyk *et al.* 1998). "National systems of innovation" have been posited for a variety of reasons, for example, because of the need to collect statistics on a national basis and in relation to national production systems (Lundvall, 1998; Nelson, 1993). In the case of Japan (Freeman, 1987), or in comparisons among Latin-American countries (Cimoli, 2000), such a delineation may provide a heuristics more than in the case of the European nations participating in the common frameworks of the European Union (Leydesdorff, 2000). Furthermore, systems of innovation may vary in terms of their strengths and weaknesses in different dimensions. For example, while one would expect a system of innovations in the Cambridge region to be highly science-based (Etzkowitz *et al.*, 2000), the system of innovations of the Basque country is industrially based and reliant on technology centers with applied research more than on universities for its knowledge base (Moso & Olazaran, 2002).

The definition of a "system of innovation" can be different from different perspectives. While the OECD, for example, focused on comparing national statistics, the EU has had a tendency to focus on changes in the interactions among the member states, for example, in trans-border regions.[1] In the meantime, Belgium has been regionalized to such an extent that one no longer expects the innovation dynamics of Flanders to be integrated with the other parts of the country. The question of which dimensions are relevant to the specificities of which innovation system requires empirical specification and research, but in order to draw conclusions from such research efforts one needs a theoretical framework. The framework should enable us to compare across innovation systems and in terms of relevant dimensions.

Three frameworks have been elaborated in innovation studies during the 1990s (Shinn, 2002):

1. the approach of comparing (national) systems of innovation (Lundvall, 1988 and 1992; Nelson, 1993; Edqvist, 1997);

2. the thesis of a new "Mode 2" in the production of scientific knowledge (Gibbons *et al.*, 1994; Nowotny *et al.*, 2001); and

3. the Triple Helix of University-Industry-Government relations (Etzkowitz & Leydesdorff, 1997, 2000; Leydesdorff & Etzkowitz, 1998).

## 3.1    National systems of innovation

Concepts like a "knowledge economy" (Machlup, 1962) and/or the "information society" (Bell, 1973; Castells, 1996) can be traced back in the literature to dates earlier than the expression "knowledge-based economy." However, the OECD (1996) launched the concept of a *knowledge-based* economy specifically in reaction to increased criticism of the *national* systems of innovation approach that had been prevalent before this date (Godin, 2004). Although one refers also to older literature (List, 1841; cf. Freeman & Soete, 1997), the "national systems of innovation" approach emerged from a fascination with the Japanese innovation system during the 1980s (Freeman, 1987, 1988; Irvine & Martin, 1984).

Lundvall (1988) elaborated the theory of "national systems of innovation" for the specific context of Scandinavia. According to this author, specificities in user-producer relations enable developers to reduce uncertainties in the market more rapidly and over longer stretches of time than in the case of less coordinated economies (Teubal, 1979).[2] Lundvall proposed that learning in interactions between users and producers—and not economic agency based on individual preferences—should be considered as the *micro-foundation* of the economy. For example, he formulated:

> When the technology changes rapidly and radically—when a new technological paradigm (for a discussion and definition, see Dosi, 1982) develops—the need for proximity in terms of geography and culture becomes even more important. A new technological paradigm will imply that established norms and standards become obsolete and that old codes of information cannot transmit the characteristics of innovative activities. In the absence of generally accepted standards and codes able to transmit information, face-to-face contact and a common cultural background might become of decisive importance for the information exchange. (Lundvall, 1988: 355).

After arguing that the interaction between users and producers belonging to the same national systems may work more efficiently for reasons of language and culture, Lundvall (1988: 360 ff.) proceeded by proposing the nation as the main system of reference for innovations. Nations can be considered as political economies constructed during the 19[th] century (List, 1841; Marx, 1848). Nowadays, they provide strong 'integrators' in otherwise differentiated economies (Nelson & Winter, 1975; Nelson, 1982; Skolnikoff, 1993). Furthermore, nation states can meaningfully be compared in terms of their innovative capacities (Lundvall, 1992; Nelson, 1993).

Within the context of the European Union, however, the concept of national systems of innovation could also be considered as defensive. A number of typically Scandinavian projects has also failed (Van den Besselaar, 1998), and it is increasingly difficult to see how Nokia—a Finnish company—embodies values that were shielded from the world market more than those of its main competitors. In another study, Edqvist (1997) generalized the concept of "systems of innovation" to cross-sectoral innovation patterns and their institutional connections (cf. Carlson & Stankiewicz, 1991; Whitley, 2001). Thus, Lundvall's argument about user-producer interactions as a micro-foundation of economic wealth production at the network level—as distinct from individual preferences—can be considered as a contribution beyond his focus on national systems.

User-producer relations contribute to a system's formation and maintainance as one of the system's subdynamics. In an early stage of the development of a technology, a close (e.g., co-evolutionary) relation between technical specifications and market characteristics can provide the development of a design with a competitive advantage (Frenken, 2001; Callon *et al.*, 2002). In a next stage, however, when the diffusion parameter becomes larger than half the internal substitution rate in the production process, one can also expect a bifurcation between these two subdynamics. The diffusion may globalize, and thereafter the globalized dimension feeds back on local production processes, for example, when a multinational corporation decides to move resources out of its country of origin.

In other words, geographical proximity can be expected to serve the incubation of new technologies. However, the regions of origin do not necessarily coincide with the contexts that profit from these technologies at a later stage of the development. Various Italian industrial districts provide examples of this flux. As companies developed a competitive edge, they may move out of the region generating a threat of deindustrialization which has to be countered at the regional level continuously (Dei Ottati, 2003; Sforzi, 2003). The four regions indicated by the EU as the 'engines of innovation' in the early 1990s were no longer the most innovative regions in the late 1990s (Boschma & Lambooy, 2002; Krauss & Wolff, 2002; Laafia, 1999; Viale & Campodall'Orto, 2002).

In summary, a localized system of innovation defined as a nation or a region can be decomposed in terms of the flows and stocks contained in this system. The institutional framework of a system of innovation has to be complemented with a functional analysis. Control—and therefore the possibility of appropriation—emerges from the recombination of institutional opportunities and functional requirements. In some cases and at certain stages of the innovation, local stabilization in a geographic proximity can be beneficial, for example, because of increased puzzle-solving capacity in a niche. However, at a subsequent stage this advantage may turn into a disadvantage because the innovations can get "locked-in" into the local conditions. Various subdynamics are competing and interacting. The result is a complex dynamics.

## 3.2    "Mode 2"

The Mode 2 thesis (Gibbons *et al.*, 1994) implies that the economic system has gained a degree of freedom under the pressure of globalization and the new communication technologies. What seemed to be rigid under a previous regime can be made flexible under the new regime of communication. In their latest book, Nowotny *et al*. (2001) specify that the new flexibility is not to be considered as a "weak contextualization," and argue that their thesis is about "strong contextualization." A system of innovation is a construct that is continuously under reconstruction and can be reconstructed even in the core of its operations. The perspectives of change are limited more in terms of codifying expectations than in terms of historical constraints. The institutions can be changed reflexively by political and technological interventions.

How does one allocate the capacities for puzzle-solving and innovation across the system? The authors formulate:

> There is no longer only one scientifically 'correct' way, if there ever was only one, especially when—as is the case, for instance, with mapping the human genome—constraints of cost-efficiency and of time limits must be taken into account. There certainly is not only one scientifically 'correct' way to discover an effective vaccine against AIDS or only one 'correct' design configuration to solve problems in a particular industry. Instead, choices emerge in the course of a project because of many different factors, scientific, economic, political and even cultural. These choices then suggest further choices in a dynamic and interactive process, opening the way for strategies of variation upon whose further development ultimately the selection through success will decide. (Nowotny *et al.*, 2001: 115f.)

The perspective, consequently, is changed from interdisciplinary to transdisciplinary. The global perspective provides us with more choices than were realized hitherto. Reflection is considered as the mechanism which can makes this difference in the discourse. Note that these authors consider reflection as a property of the communication. The communication adds another dimension to the reflection by individual agents. In other words, while Lundvall (1988) focused already on interaction and argued that communications can stabilize the local innovation environment, these authors argue that communications enable us to entertain a global perspective on the relevant environments. Communications can develop an internal dynamics.

Reflection provides meaning from a hindsight perspective. Thus, the reflexive perspective adds a dynamic that is different from the historical one. While the latter focuses on the opportunities and constraints of a given unit (e.g., a region), the discourse enables us to redefine a system of reference by contextualizing and analyzing it. Therefore, the focus shifts methodologically from the historical reconstruction of a system by "following the actors" (Latour, 1987) to the functional analysis of a hypothesized (!) innovation system in the present. The robustness of the construct depends not only on its intrinsic quality, but on the level of support that can be mobilized from other subsystems of society (e.g., the economy or the political systems involved).

What does this model add to the model of "national innovation systems" in terms of the micro-foundation? The micro-economics of Lundvall was grounded no longer in terms of the preferences of agents, but in terms of communication and interaction between users and producers. The authors of "Mode 2" define another communication dynamic relevant to the systems of innovation. While agency can be considered as a source of communication—and can be expected to be reflexive, for example, in entertaining preferences—it has necessarily a position. Communication, however, is relational and develops among communicating agents in the present with an emergent dynamics.

The links of a communication system can be expected to operate differently from the nodes. Categories like reflexivity and knowledge may have a different meaning with reference to the one or the other layer of the system. For example, the structure of communication exhibits preferences by providing certain agents with access and others not. In addition to actions which provide the variation, the internal dynamics of communications are able to generate change at the systemic level, that is, in terms of the structural selections. Although the communication dynamics can be recognized by the actors, the network dynamics can be expected to remain partially latent (Lazarsfeld & Henry, 1968; Giddens, 1981).

Luhmann (1984) has proposed that communication be considered as a system of reference different from agency. He emphasized the analytical independence of this perspective (e.g., Luhmann, 1996). The two systems of agency and communication are considered as "structurally coupled" in the action as an event (Maturana, 1978). This event can be attributed as an action to the actor, while it can be expected to function as a communication within the communication system.

Social systems communicate reflexively in addition to communicating in terms of first-order exchange relations. The relations develop along the time axis, but meaning is provided to the exchanges from a perspective of hindsight. Furthermore, various meanings can be codified differently. For example, in scientific communications "energy" has a meaning different from its meaning in the political discourse. While economists and politicians are able to worry about "shortages of energy," "energy" is defined as a conserved quantity in physics. Words may have different meanings in other contexts. Thus, the evolutionary dynamics of social communication adds another layer of complexity to the first-order dynamics of the exchange. The institutionalization and organization stabilize the communication historically, but by providing meaning to the communication one potentially generates a set of global perspectives (Urry, 2003).

Global perspectives can be focused when the communications are specifically codified. For example, scientific communications enable us to deconstruct and reconstruct phenomena in great detail. As noted above, the price mechanism can be further refined in terms of price/performance ratios. Political communication is channeled in a democracy through the *trias politica* (Montesquieu, 1748). The differentiation of the communication into various functions enables the social system to process more complexity than in a hierarchical mode. However, under this condition one can expect to lose a central point of coordination and control as the interacting (sub)systems of communication become increasingly self-organized. This communication regime reshapes the existing communication structures as in a cultural evolution. Mechanisms of selection other than "natural" ones reconstruct the system from perspectives potentially different from those that were historically available (Leydesdorff, 2003a).

In summary, the communicative layer provides society with a selection environment for historical institutions. While variation can be expected to contain randomness, selection is deterministic. However, the communication structures of the social system are complex because the codes of the communication have been differentiated. Communications develop along the various axes, but they can additionally be translated by using different codes at the interfaces reflexively.

Interaction terms among codes of communication emerged as a matter of concern within knowledge-based corporations when interfaces between R&D and marketing had increasingly to be managed (Galbraith, 1967). In university-industry-government relations, however, three types of communications are interfaced. In the frictions between the institutional layer of these relations and the dynamics of the mutual expectations, the system can be expected to gain another degree of freedom. The utilization of this degree of freedom increasingly provides a competitive advantage.

## 3.3    The Triple Helix

While the systems-of-innovation approach defined innovation systems exclusively in terms of (aggregates of) institutional units of analysis, 'Mode 2' defined innovations exclusively in terms of reconstructions on the basis of emerging perspectives in the communication. The Triple Helix approach combines these two perspectives as different subdynamics of the systems under study. However, this model adds the dynamics of the market as its third perspective with a micro-foundation in neo-classical economics. Thus, one can assume that an innovation system can be driven by various subdynamics to variable extents. Consequently, the discussion shifts from an ontological one about what an innovation system 'is' to the methodological question of how one can study innovation systems in terms of these potentially different dimensions.

In the Triple Helix model, the main institutions carrying the knowledge-based system are first analyzed as university, industry, and government. These carriers of an innovation system entertain a dually layered network: one layer of institutional relations in which they constrain each other's behaviour, and another one of functional relations in which they shape each other's expectations. The functions that have to be recombined and reproduced evolutionarily can be specified as (a) wealth generation in the economy, (b) novelty generation by organized science and technology, and (c) control of these two functions locally for the retention and reproduction of the system. The layers can be expected to feed back on each other, thus changing the institutional roles, the institutional environments, and therefore potentially the evolutionary functions of the various stakeholders in a next round.

Within this complex dynamic, the two mechanisms specified above—'user-producer relations' and reflexive communication, respectively—can be considered as additional to the micro-foundation in neo-classical economics. First, each agent or aggregate of agencies is positioned differently in terms of preferences and other attributes. Secondly, the agents interact, for example, in economic exchange relations. This generates the networks. Thirdly, the arrangements of positions (nodes) and relations (links) can be expected to contain information because not all network positions are held equally and links are selectively generated and maintained.

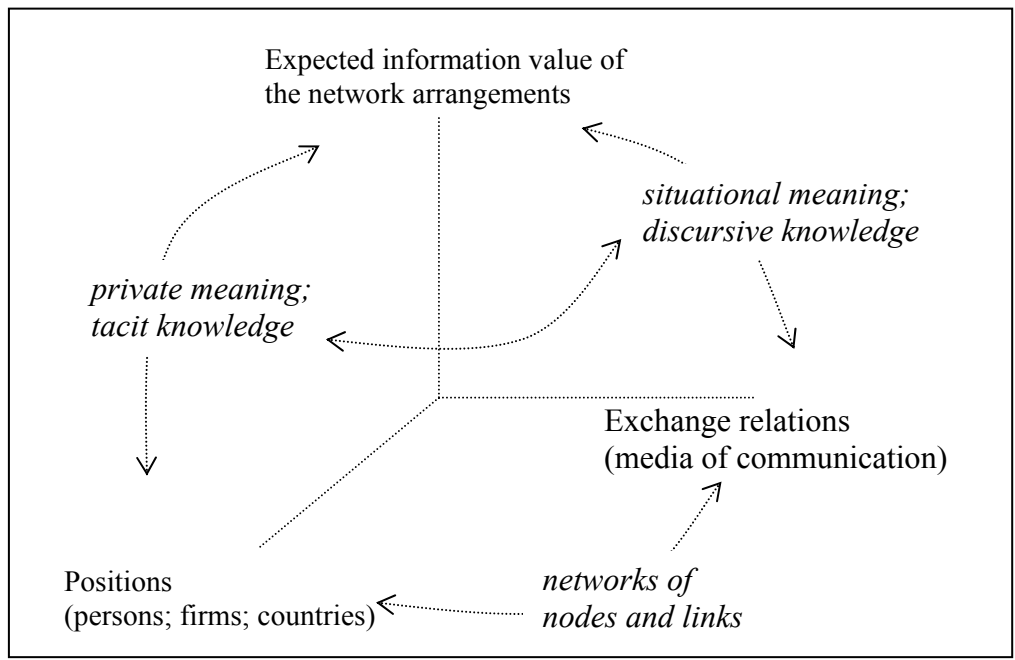

**Figure 1**

Micro-foundation of the Triple Helix Model of Innovation

The expected information content of the distributions can be *recognized* by relevant agents at local nodes. This recognition generates knowledge within these agents and their organizations. Knowledge, however, can also be processed as discursive knowledge in the network of relations. Knowledge that is communicated can be further codified by the sciences (Cowan & Foray, 1997). Figure 1 summarizes the configuration.

If one generalizes this model to the level of a social system, three analytically independent dimensions of an innovation system can be distinguished (Figure 2): (1) the geography which organizes the positions of agents and their aggregates; (2) the economy organizing the exchange relations; and (3) the knowledge content which emerges with reference to either of these dimensions (Archer, 1995). Given these specifications, one can model the relevant dimensions and their interaction terms as follows:

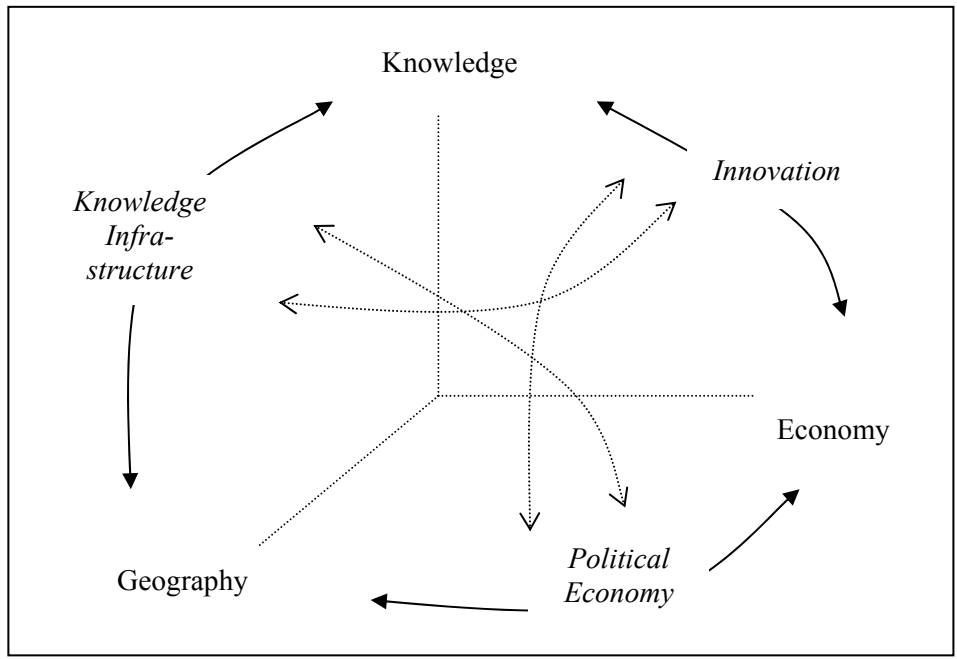

**Figure 2**

Three dimensions of an innovation system with the first-order interaction terms

The knowledge base of an economy can be considered as a second-order interaction effect between the first-order interaction effects depicted in Figure 2. As new solutions are found for interfacing the three components, new technological trajectories can be shaped at interfaces along the time axis (Dosi, 1982; Frenken & Leydesdorff, 2000). Technological trajectories develop with the arrow of time. However, technological regimes and paradigms feed back as next-order systems in terms of expectations that emerge with hindsight (Leydesdorff & Van den Besselaar, 1998). In other words, a technological regime remains pending as selection pressure on the technological trajectories by which it is generated and reproduced.

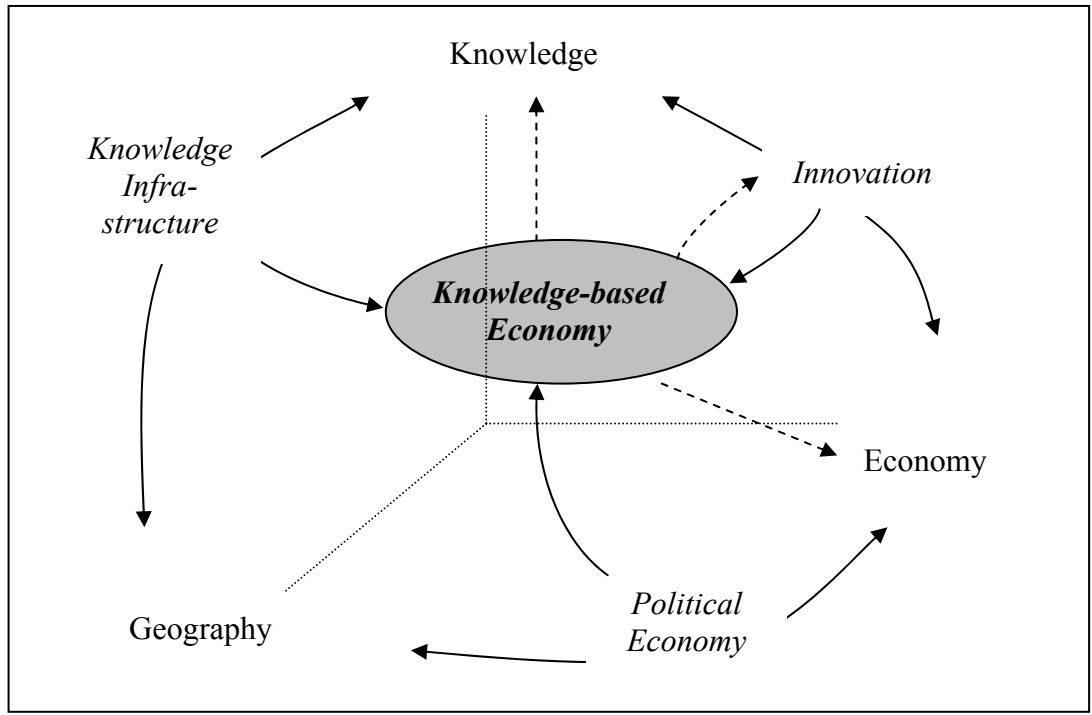

**Figure 3**
The first-order interactions generate a knowledge-based economy as a next-order system (from: Leydesdorff & Meyer, 2003).

Figure 3 visualizes the knowledge base of an economy as this second-order interaction effect among the first-order interactions at interfaces. A second-order interaction effect cannot occur historically before the first-order interaction effects are firmly in place. Upon its occurrence, however, the emerging resonance may work as an attractor that entrains and therefore reorganizes the fluxes of communications on which it rests (Kampmann *et al.*, 1994).

A niche of knowledge-based self-organization can be maintained locally as a pocket of negative entropy. The reduction of uncertainty is brought about by the expectation of a next-order regime. Note that this expectation is based on reflexive interactions. However, the localizing anchor of such a system is no longer given. It can be positioned in the geographical dimension and/or in terms of sectors within an economy and/or in terms of a techno-science as in the case of biotechnology (Carlsson, 2002). When the structural interaction terms are changed for historical reasons (as in Eastern Europe during the 1990s), the second-order interaction terms can be expected to change. The respective systems of innovation may then have entered into another basin of attraction.

## 4.      Structural change

Before turning to the issue of how the knowledge base of an economy can be indicated as a next-order interaction among three two-dimensional interfaces, let me first specify the mechanisms of structural change within the self-organization of a complex system. As noted, one expects interaction terms among both the functions and the institutional carriers of these functions. The layers of functionality and institutionalization can couple either vertically (as when there is, for example, one-to-one correspondence between functions and the carrying institutions) or horizontally (as in university-industry-government relations). The horizontal coupling among institutions could be specified above as containing a dynamic different from the horizontal coupling among functions. Functions couple evolutionarily (as an anticipatory feedback), while institutions couple historically, that is, with the arrow of time (Coveney & Highfield, 1990).

The disturbance terms in the couplings provide the variation. However, the subsystems also select upon one another while being developed recursively. The disturbances are needed for the further development of "requisite variety" (Ashby, 1958) in innovation systems. Because of the disturbances, the systems cannot be completed in one direction or another. Innovation systems, therefore, can be expected to remain in transition (Etzkowitz & Leydesdorff, 1998). However, the various subdynamics mainly operate as selection mechanisms upon one another. Selection is a recursive operation: some selections can be selected for stabilization, and some stabilizations can be selected for globalization. This recursivity in the selection may easily extinguish the signal and lead to relative stagnation. Resilience to innovation can hence be considered as the default of a complex system despite its capacity to develop new variants at lower levels continuously (Bruckner *et al.,* 1994).

Under which conditions can the variation—that is produced as disturbance because of the ongoing interactions—lead to structural change at the system's level? Two mechanisms are important for understanding structural change as a mechanism despite the resilience prevailing globally: (1) the possibility of "lock-in" between subsystems, and (2) the possibility of bifurcation within an evolving system. A bifurcation generates an interface (for example, between a production and a diffusion system), while a "lock-in" tends to dissolve an interface in a coevolution, for example, between a specific technology and a market. The locked-in systems may bifurcate in a next stage. However, the two mechanisms are analytically independent.

For example, a computer shop can be expected to replenish its supplies when it sells specific brands more than others. The sales patterns of the previous period can be reinforced by the adjusted offering at a next moment in time. Over time, this leads to a profile corresponding to the market segments that are locally most attractive. When supply and demand match, the system of two subsystems (technology and market) may become locked-in. However, when demand increases further, the market may be segmented or otherwise differentiated. The stability gained previously meta-stabilizes in this case and then potentially globalizes.[3]

"Lock-in" was modeled by Arthur (1989, 1994) in terms of the Polýa urn model, but the principle is better known through the historical work of David (1985) on the QWERTY keyboard (cf. Liebowitz & Margolis, 1999). One can be locked-in into suboptimal technologies when the volume of a specific technology has generated network effects that feedback on the adoption of this technology. Instead of decreasing marginal returns, the economies of scale can then be characterized as leading to increasing marginal returns. Information technologies exhibit this type of economy to a larger extent than the industrial technologies of the previous period (Arthur *et al.*, 1997). In the case of "lock-in," however, the system loses a degree of freedom because of a tight coupling between dimensions in a coevolution. For example, the choice among technologies may no longer be available.

The reaction-diffusion mechanism makes it possible to gain degrees of freedom at interfaces. In an early stage of the development of a new technology, the processes of invention, production, and marketing can be small-scale and intensively coupled, for example, within a corporation or a geographical region. However, the diffusion mechanism contains a dynamic other than the production process. It can be shown that the two coupled mechanisms will bifurcate when the diffusion parameter becomes larger than half of the rate in the production process (Turing, 1952; cf. Rosen, 1985: 183f.).[4] This bifurcation can be expected to lead to an uncoupling of the hitherto tightly coupled systems. Thus, a degree of freedom can be gained at the systemic level. Once the diffusion is uncoupled from the production, the former may feedback on the local production, for example, when the internationalization leads to disinvestments in the country of origin.

In summary, the two evolutionary mechanisms of lock-in and bifurcation provide us with a mechanism for structural change within innovation systems. On the one hand, lock-in leads to a new dimension along which a new trajectory can be expected to develop, but at the price of sacrificing a degree of freedom. Bifurcation, on the other hand, provides a mechanism to regain a degree of freedom, but with reference to the newly produced dimension. Via the emergence of a new trajectory, the system can move from one basin of attraction to another.

## 5. Indicators of the knowledge base of an economy

In an evolutionary model, the institutional dimensions can no longer be expected to correspond one-to-one to the functions carried by and among the agencies. By using the Triple Helix model, however, one can identify in a first-order approximation the university as the main carrier of the knowledge production function, industry as a proxy of the economic function, and the relevant level of government as responsible for the interfacing and organization of the two other functions at the respective systems level. The three-dimensional system can be operationalized by measuring the variation in each dimension and the co-variations between them. One then obtains the following scheme of university-industry-government (UIG) relations:

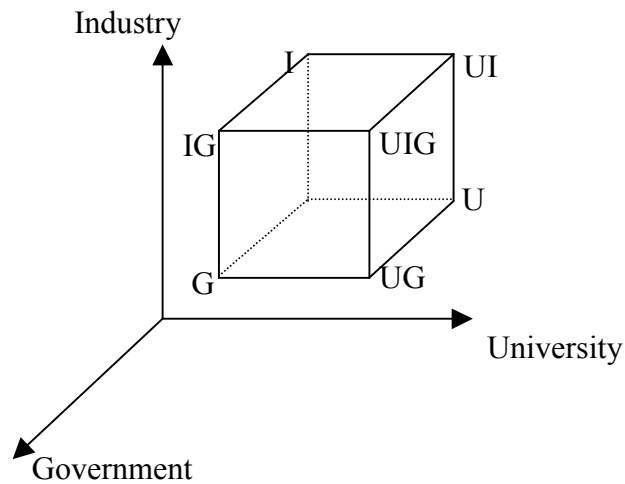

**Figure 4**

The three-dimensions of measurement in a Triple Helix configuration and their combinations

As noted, the knowledge base of an economy can be considered as a second-order interaction effect among the various bilateral relations. Unlike the co-variation between two dimensions the mutual information in three dimensions can be negative. This potentially negative entropy provides us with an indicator for the self-organization in the arrangements among the fluxes of communication in institutional arrangements (Leydesdorff, 2003b).

It can be shown that the mutual information in a relation between two systems is always positive or zero (in the case of no relation), but this probabilistic entropy cannot be negative (Theil, 1972). The mutual information in three dimensions, however, can become negative if the bilateral relations form a network system (Abramson, 1963: 129 ff.). The dynamics of the system can increasingly be determined by the communication links among the subsystems. Note that such a system is no longer agent-based at the nodes, but communication-based in terms of the network of relations. In other words, the overlay system may function increasingly as another subdynamic that reduces the uncertainty that prevails at the systemic level.

How can a negative entropy produced at the network level reduce the uncertainty within a three-dimensional system? For example, in a family a child can entertain relations with both its parents. But in addition to these two components of the expected information content of this family system and its common component in the aggregate (when the three are together, for example, at dinner time), the relation between the two parents can reduce the uncertainty for the child beyond its control. Thus, the latent structure of relations sometimes reduces the uncertainty locally as a feedback mechanism on the historically unfolding events.

This degree of freedom is declared in the Triple Helix model as an overlay of expectations that continuously restructures the underlying arrangements, but to a variable extent. The mutual information in three dimensions can be used as an indicator of this operation because it indicates the transmission at the network level (Theil, 1972; Leydesdorff, 1995). The formalization of the measure is as follows:

$$T_{UIG} = H_U + H_I + H_G - H_{UI} - H_{IG} - H_{UG} + H_{UIG} \qquad (1)$$

While each of the interacting systems ($H_U$, $H_I$, and $H_G$) adds to the uncertainty, the uncertainty at the system's level is reduced by the mutual relations at the interfaces between them. As in the two-dimensional case, the systems condition and determine each other mutually in their relations. The three-dimensional uncertainty in the overlap ($H_{UIG}$), however, adds again positively to the uncertainty that prevails at the network level. Because of the alteration in the signs, the three-dimensional transmission can become negative when the reduction of the uncertainty by the bi-lateral relations prevails over the central coordination. Note that the value of $T_{UIG}$ can be calculated directly from the relative frequency distributions of the variables and their co-variation by using the Shannon-formulas.[5] What is to be considered as the "university" system, the "industry" system or the "government" system, however, has to be specified on theoretical grounds.

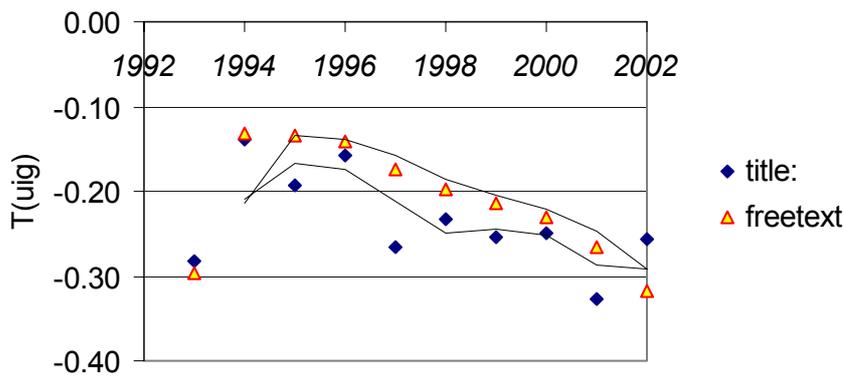

**Figure 5**

The mutual information among 'university,' 'industry,' and 'government' relations as retrieved at the Internet using the *AltaVista Advanced Search Engine*. (The curves are based on two-year moving averages. The measurements were performed on 15 May 2003).

In another context I have elaborated several examples of the measurement of this mutual information in three dimensions using scientometric, webometric, and U.S. patent data (Leydesdorff, 2003b). Figure 5 provides one example of the co-occurrences of the words 'university,' 'industry,' and 'government' retrieved using the AltaVista Advanced Search Engine for different years. The development of this data over time shows the increasingly negative value of the mutual information among these three dimensions over time. The pattern follows the globalization of the Internet after its initial phase of 1993-1995.[6]

## 6.    Conclusions

The Triple Helix model can be elaborated as a neo-evolutionary model that enables us first to understand innovation systems as complex systems and then also serves us as a heuristic for the operationalization. A micro-foundation of the model could be proposed by arguing that the knowledge base of an economy is not only agent-based, but also and in important respects communication-based. When information is exchanged among agents, the communication can be provided with meaning, and this meaning can be further codified, for example, by developing knowledge. Knowledge can perhaps be considered as a meaning that makes a difference. Knowledge informs the exchanges and thus selectively reinforces the solutions found hitherto at interfaces. This latter operation potentially reduces the uncertainty, lowers transaction costs, and thus transforms interfaces within the system innovatively. As knowledge is further organized and codified, for example, in terms of science and technology, this "globalizing"—that is, meta-stabilizing—dynamic of an economy can be reinforced.

The new dynamic of a knowledge-based economy has important consequences for the function of regions. The locales may serve as the incubators where production, innovation, and diffusion processes are closely coupled. The density of the local interactions increases the chances for "lock-in" and therefore the (co-)shaping of trajectories within the system (Callon *et al.*, 2002). The density of the interactions within clusters and regions determines this capacity. Therefore, one can expect metropolitan regions to hold an advantageous position in the new economy.

The development of innovative interfaces, however, is not a sufficient condition for success in a knowledge-based configuration. Second-order interactions among the interfaces are a condition for the reduction of uncertainty at the system's level. Beyond the two-dimensional interfaces a three-dimensional system can indicate this reduction of the uncertainty. The development of a three-dimensional indicator was proposed above as an instrument to orient our systems of innovation on the ocean provided by a knowledge-based economy.

## References

Abernathy, W. J. and J. Utterback. 1978. Patterns of Industrial Innovation. *Technology Review*, 50, 41-47.

Abramowitz, M. 1956. Resource and Output Trends in the United States since 1870. *American Economic Review,* 46, 5-23.

Abramowitz, M. and P. A. David. 1996. Measuring Performance of Knowledge-Based Economy. In *Employment and Growth in the Knowledge-Based Economy* (pp. 35-60). Paris: OECD.

Archer, M. S. 1995. *Realist Social Theory: The Morphogenetic Approach*. Cambridge, UK: Cambridge University Press.

Arrow, K. J. 1962. The Economic Implications of Learning by Doing. *Review of Economic Studies*, 29, 155-173.

Arthur, W. B. 1989. Competing Technologies, Increasing Returns, and Lock-in by Historical Events. *Economic Journal*, 99, 116-131.

Arthur, W. B. 1994. *Increasing Returns and Path Dependence in the Economy*. Ann Arbor: University of Michigan Press.

Arthur, W. B., S. N. Durlauf and D. A. Lane (Eds.). 1997. *The Economy as an Evolving Complex System II*. Redwood, CA: Addison-Wesley.

Ashby, R. 1958. Requisite Variety and Its Implications for the Control of Complex Systems. *Cybernetica*, 1 (2), 1-17.

Barras, R. 1990. Interactive Innovation in Financial and Business Services: The Vanguard of the Service Revolution. *Research Policy*, 19, 215-237.

Beccatini, G., M. Bellandi, G. Dei Ottati and F. Sforzi. 2003. *From Industrial Districts to Local Development: An Itinerary of Research*. Centenham, UK; Norhhampton, MA: Edward Elgar.

Bell, D. 1973. *The Coming of the Post-Industrial Society*. New York: Basic Books.

Biggiero, L. 1998. Italian Industrial Districts: A Triple Helix Pattern of Problem Solving. *Industry and Higher Eductation*, 12 (4), 227-234.

Boschma, R. A. and J. G. Lambooy. 2002. Knowledge, Market Structure and Economic Coordination: The Dynamics of Italian Industrial Districts. *Growth and Change,* 33 (3), 291-311.

Braczyk, H.-J., P. Cooke, and M. Heidenreich (Eds.) 1998. *Regional Innovation Systems*. London/ Bristol PA: University College London Press.

Bruckner, E., W. Ebeling, M. A. J. Montaño, and A. Scharnhorst. 1994. Hyperselection and Innovation Described by a Stochastic Model of Technological Evolution. In L. Leydesdorff & P. v. d. Besselaar (Eds.), *Evolutionary Economics and Chaos Theory: New Directions in Technology Studies* (pp. 79-90). London: Pinter.

Callon, M. 1998. *The Laws of the Market.* Oxford and Malden, MA: Blackwell.

Callon, M., C. Méadel, and V. Rabeharisoa. 2002. The Economy of Qualities. *Economy and Society*, 31 (2), 194-217.

Carlsson, B. (Ed.) 2002. *New Technological Systems in the Bio Industries -- an International Study*. Boston/Dordrecht/London: Kluwer Academic Publishers.

Carlsson, B. and R. Stankiewicz. 1991. On the Nature, Function, and Composition of Technological Systems. *Journal of Evolutionary Economics,* 1 (2), 93-118.

Castells, M. 1996. *The Rise of the Network Society.* Oxford and Malden, MA: Blackwell.

Cimoli, M. (Ed.). 2000. *Developing Innovation Systems: Mexico in a Global Context.* London: Continuum.

Council of the European Communities, and Commission of the European Communities. 1992. *Treaty on European Union.* Luxemburg: Office for Official Publications of the European Communities.

Coveney, P. and R. Highfield. 1990. *The Arrow of Time*. London: Allen.

Cowan, R. and D. Foray 1997. The Economics of Codification and the Diffusion of Knowledge. *Industrial and Corporate Change,* 6, 595-622.

David, P. A. 1985. Clio and the Economics of Qwerty. *American Economic Review*, 75, 332-337.

Dei Ottati, G. 2003. Local Governance and Industrial Districts' Competitive Advantage. In G. Beccatini, M. Bellandi, G. D. Ottati, and F. Sforzi (Eds.), *From Industrial Districts to Local Development: An Itinerary of Research* (pp. 108-130). Cheltenham, UK/ Northhampton, MA: Edward Elgar.

Dosi, G. 1982. Technological Paradigms and Technological Trajectories: A Suggested Interpretation of the Determinants and Directions of Technical Change. *Research Policy*, 11, 147-162.

Edqvist, C. (Ed.) 1997. *Systems of Innovation: Technologies, Institutions and Organizations.* London: Pinter.

Etzkowitz, H. and L. Leydesdorff. 1998. The Endless Transition: A "Triple Helix" of University-Industry-Government Relations. *Minerva*, 36, 203-208.

Etzkowitz, H. and L. Leydesdorff. 2000. The Dynamics of Innovation: From National Systems and 'Mode 2' to a Triple Helix of University-Industry-Government Relations. *Research Policy*, 29 (2), 109-123.

Etzkowitz, H. and L. Leydesdorff (Eds.). 1997. *Universities in the Global Knowledge Economy: A Triple Helix of University-Industry-Government Relations.* London: Pinter.

Etzkowitz, H., A. Webster, C. Gebhardt, and B. R. C. Terra. 2000. The Future of the University and the University of the Future: Evolution of Ivory Tower to Entrepreneurial Paradigm. *Research Policy*, 29 (2), 313-330.

Foray, D. and B.-A. Lundvall. 1996. The Knowledge-Based Economy: From the Economics of Knowledge to the Learning Economy. In *OECD Documents: Employment and Growth in the Knowledge-Based Economy* (pp. 11-32). Paris: OECD.

Freeman, C. 1987. *Technology Policy and Economic Performance: Lessons from Japan*. London: Pinter.

Freeman, C. 1988. Japan, a New System of Innovation. In G. Dosi, C. Freeman, R. R. Nelson, G. Silverberg & L. Soete (Eds.), *Technical Change and Economic Theory* (pp. 31-54). London: Pinter.

Freeman, C. and C. Perez. 1988. Structural Crises of Adjustment, Business Cycles and Investment Behaviour. In G. Dosi, C. Freeman, R. N. G. Silverberg, and L. Soete (Eds.), *Technical Change and Economic Theory* (pp. 38-66). London: Pinter.

Frenken, K. 2001. *Understanding Product Innovation Using Complex Systems Theory*. Unpublished Ph. D. Thesis, University of Amsterdam, Amsterdam.

Frenken, K. and L. Leydesdorff. 2000. Scaling Trajectories in Civil Aircraft (1913-1970). *Research Policy*, 29 (3), 331-348.

Galbraith, J. K. 1967. *The New Industrial State*. Penguin: Harmondsworth.

Gibbons, M., C. Limoges, H. Nowotny, S. Schwartzman, P. Scott, and M. Trow. 1994. *The New Production of Knowledge: The Dynamics of Science and Research in Contemporary Societies*. London: Sage.

Giddens, A. 1981. Agency, Institution, and Time-Space Analysis. In K. D. Knorr-Cetina & A. V. Cicourel (Eds.), *Advances in Social Theory and Methodology. Toward an Integration of Micro- and Macro-Sociologies* (pp. 161-174). London: Routledge & Kegan Paul.

Godin, B. 2004. The Knowledge-Based Economy: Conceptual Framework or Buzzword, *Journal of Technology Transfer* (this issue).

Hall, P. A. and D. W. Soskice (Eds.). 2001. *Varieties of Capitalism: The Institutional Foundations of Comparative Advantage*. Oxford, etc.: Oxford University Press.

Hughes, T. P. 1987. The Evolution of Large Technological Systems. In W. Bijker, T. P. Hughes & T. Pinch (Eds.), *The Social Construction of Technological Systems* (pp. 51-82). Cambridge, MA: MIT Press.

Irvine, J. and B. R. Martin. 1984. *Foresight in Science: Picking the Winners*. London: Frances Pinter.

Kampmann, C., C. Haxholdt, E. Mosekilde, and J. D. Sterman. 1994. Entrainment in a Disaggregated Long-Wave Model. In L. Leydesdorff & P. v. d. Besselaar (Eds.), *Evolutionary Economics and Chaos Theory: New Directions in Technology Studies* (pp. 109-124). London/New York: Pinter.

Kemp, R., J. Schot, and R. Hoogma. 1998. Regime Shifts to Sustainability through Processes of Niche Formation. The Approach of Strategic Niche Management. *Technology Analysis and Strategic Management*, 10 (2), 175-195.

Krauss, G. and H.-G. Wolff. 2002. Technological Strengths in Mature Sectors--an Impediment of an Asset of Regional Economic Restructuring? The Case of Multimedia and Biotechnology in Baden-Württemberg. *Journal of Technology Transfer,* 27 (1), 39-50.

Laafia, I. 1999. *Regional Employment in High Technology*. Eurostat, at http://europa.eu.int/comm/eurostat

Lancaster, K. J. 1979. *Variety, Equity and Efficiency*. New York: Columbia University Press.

Latour, B. 1987. *Science in Action*. Milton Keynes: Open University Press.

Leydesdorff, L. 2000. Is the European Union Becoming a Single Publication System? *Scientometrics*, 47 (2), 265-280.

Leydesdorff, L. 2001. Indicators of Innovation in a Knowledge-Based Economy. *Cybermetrics*, 5 (Issue 1), Paper 2, at http://www.cindoc.csic.es/cybermetrics/articles/v5i1p2.html .

Leydesdorff, L. 2003a. The Construction and Globalization of the Knowledge Base in Inter-Human Communication Systems. *Canadian Journal of Communication*, 28 (3), 267-289.

Leydesdorff, L. 2003b. The Mutual Information of University-Industry-Government Relations: An Indicator of the Triple Helix Dynamics. *Scientometrics*, 58 (2), 445-467.

Leydesdorff, L. and P. v. d. Besselaar. 1998. Technological Development and Factor Substitution in a Non-Linear Model. *Journal of Social and Evolutionary Systems*, 21, 173-192.

Leydesdorff, L. and H. Etzkowitz. 1998. The Triple Helix as a Model for Innovation Studies. *Science and Public Policy*, 25 (3), 195-203.

Leydesdorff, L., P. Cooke, and M. Olazaran (Eds.) 2002. Regional Innovation Systems in Europe (Special Issue). *Journal of Technology Transfer*, 27 (1), 5-145.

Liebowitz, S. J. and S. E. Margolis 1999. *Winners, Losers & Microsoft: Competition and Antitrust in High Technology*. Oakland, CA: The Independent Institute.

List, F. 1841. *The National Systems of Political Economy*. London: Longman, 1904.

Luhmann, N. 1996. On the Scientific Context of the Concept of Communication. *Social Science Information*, 35 (2), 257-267.

Lundvall, B.-Å. 1988. Innovation as an Interactive Process: From User-Producer Interaction to the National System of Innovation. In G. Dosi, C. Freeman, R. Nelson, G. Silverberg & L. Soete (Eds.), *Technical Change and Economic Theory* (pp. 349-369). London: Pinter.

Lundvall, B.-Å. (Ed.). 1992. *National Systems of Innovation*. London: Pinter.

Machlup, F. 1962. *The Production and Distribution of Knowledge in the United States*. Princeton: Princeton University Press.

Marx, K. 1848. *The Communist Manifesto*. Paris. (Translated by Samuel Moore in 1888.) Harmondsworth: Penguin, 1967.

Montesquieu, C. de Sécondat, Baron de (1748. *De l'esprit des lois*. Paris.

Moso, M. and M. Olazaran. 2002. Regional Technology Policy and the Emergence of an R&D System in the Basque Country. *Journal of Technology Transfer,* 27 (1), 61-75.

National Science Board. 1987. *Science & Engineering Indicators - 1987*. Washington, DC: National Science Foundation.

Nelson, R. R. (Ed.). 1982. *Government and Technical Progress: A Cross-Industry Analysis*. New York: Pergamon.

Nelson, R. R. (Ed.). 1993. *National Innovation Systems: A Comparative Analysis*. New York: Oxford University Press.

Nelson, R. R. and S. G. Winter. 1975. Growth Theory from an Evolutionary Perspective: The Differential Productivity Growth Puzzle. *American Economic Review*, 65, 338-344.

Nelson, R. R. and S. G. Winter. 1977. In Search of Useful Theory of Innovation. *Research Policy*, 6, 35-76.

Neurath, Otto, Rudolf Carnap, and Hans Hahn. 1929. *Wissenschaftliche Weltauffassung — Der Wiener Kreis*. Vienna: Veröffentlichungen des Vereins Ernst Mach.

Nowotny, H., P. Scott and M. Gibbons. 2001. *Re-Thinking Science: Knowledge and the Public in an Age of Uncertainty*. Cambridge, etc: Polity.

OECD 1980. *Technical Change and Economic Policy*. Paris: OECD.

OECD 1996. *OECD Economic Outlook, No. 60*. Paris: OECD.

OECD/Eurostat. 1997. *Proposed Guidelines for Collecting and Interpreting Innovation Data, "Oslo Manual"*. Paris: OECD.

Pavitt, K. 1984. Sectoral Patterns of Technical Change: Towards a Theory and a Taxonomy. *Research Policy*, 13, 343-373.

Porter, M. E. 1990. *The Competitive Advantage of Nations*. London: Macmillan.

Rosenberg, N. 1982. Learning by Using. In *Inside the Black Box: Technology and Economics* (pp. 120-140). Cambridge, etc.: Cambridge University Press.

Rothwell, R. and W. Zegveld. 1981. *Industrial Innovation and Public Policy*. London: Pinter.

Rousseau, R. 1999. Daily Time Series of Common Single Word Searches in Altavista and Northernlight. *Cybermetrics* 2/3, Paper 2 at http://www.cindoc.csic.es/cybermetrics/articles/v2i1p2.html

Sahal, D. 1985. Technological Guideposts and Innovation Avenues. *Research Policy,* 14, 61-82.

Saviotti, P. P. 1996. *Technological Evolution, Variety and the Economy.* Cheltenham & Brookfield: Edward Elgar.

Sforzi, F. 2003. The 'Tuscan Model' and Recent Trends. In G. Beccatini, M. Bellandi, G. dei Ottati, and F. Sforzi (Eds.), *From Industrial Districts to Local Developments: An Itinerary of Research* (pp. 29-61). Cheltenham, UK/ Northhampton, MA: Edward Elgar.

Shinn, T. 2002. The Triple Helix and New Production of Knowledge: Prepackaged Thinking on Science and Technology. *Social Studies of Science*, 32 (4), 599-614.

Skolnikoff, E. B. 1993. *The Elusive Transformation: Science, Technology and the Evolution of International Politics*. Princeton, NJ: Princeton University Press.

Solow, R. M. 1957. Technical Progress and the Aggregate Production Function. *Review of Economics and Statistics*, 39, 312-320.

Teubal, M. 1979. On User Needs and Need Determination. Aspects of a Theory of Technological Innovation. In M. J. Baker (Ed.), *Industrial Innovation. Technology, Policy and Diffusion* (pp. 266-289). London: Macmillan Press.

Turing, A. M. 1952. *Philos. Trans. R. Soc. B.*, 237, 5-72.

Urry, J. 2003. *Global Complexity*. Cambridge, UK: Polity.

Utterback, J. and F. F. Suarez. 1993. Innovations, Competition, and Industry Structure. *Research Policy*, 22, 1-22.

Van den Besselaar, P. 1998. Technology and democracy, the limits to steering. In R. H. Chatfield, S. Kuhn, M. Muller (Eds.), *Broadening Participation - 5th PDC* (pp. 1-10). Seattle: CPSR.

Viale, R. and S. Campodall'Orto. 2002. An Evolutionary Triple Helix to Strengthen Academy-Industry Relations: Suggestions from European Regions. *Science and Public Policy,* 29 (3), 154-168.

Whitley, R. D. 1999. *Divergent Capitalisms: The Social Structuring and Change of Business Systems.* New York: Oxford University Press.

Whitley, R. D. 2001. National Innovation Systems. In N. J. Smelser & P. B. Baltes (Eds.), *International Encyclopedia of the Social and Behavioral Sciences* (pp. 10303-10309). Oxford, UK: Elsevier.

Williamson, O. 1985. *The Economic Institutions of Capitalism*. New York: Free Press.

## Notes

[1] The Maastricht Treaty (1991) assigned an advisory role to the European Committee of Regions (Council of the European Communities, 1992). This role was further strengthened by the Treaty of Amsterdam in 1997, which envisaged direct consultations between this Committee of Regions and the European Parliament.

[2] The literature is also known as the "varieties of capitalism" debate (Hall & Soskice, 2001): while the Anglo-Saxon model can be characterized as a liberal market economy, coordinated market economies may provide competitive advantages by including more elements of consultation and of the welfare state (Whitley, 1999).

[3] From an historical perspective (that is, with the arrow of time) one would consider the bifurcation as a destabilization, but from an evolutionary perspective (that is, with hindsight) one can consider it as a meta-stabilization because a degree of freedom is gained. Initially, a bifurcation is fixed as an oscillation, but when the frequency interacts with other bifurcations in the complex system, a more complex dynamics (including the generation of negative entropy in the mutual relations in three or more dimensions) can be expected. Thus, the degree of freedom in the meta-stabilization can be used by a system for its globalization.

[4] Technically, in this case the system matrix of the two equations contains two eigenvalues with opposite signs, and consequently the steady state becomes a saddle point. The two trajectories are then expected to diverge.

[5] $H_i = -\sum p_i \log p_i$, $H_{ij} = -\sum p_{ij} \log p_{ij}$, etc.

[6] The AltaVista search engine provides a reconstruction of the development from a hindsight perspective while it is continuously developing (Leydesdorff, 2001). During the measurement, I controlled for instabilities in the recall (Rousseau, 1999).